\documentstyle[galley,graphicx]{mn}

\newif\ifAMStwofonts

\ifoldfss
  \ifCUPmtlplainloaded \else
    \NewTextAlphabet{textbfit} {cmbxti10} {}
    \NewTextAlphabet{textbfss} {cmssbx10} {}
    \NewMathAlphabet{mathbfit} {cmbxti10} {} 
    \NewMathAlphabet{mathbfss} {cmssbx10} {} 
  \fi
  \ifAMStwofonts
    \ifCUPmtlplainloaded \else
      \NewSymbolFont{upmath} {eurm10}
      \NewSymbolFont{AMSa} {msam10}
      \NewMathSymbol{\upi}     {0}{upmath}{19}
      \NewMathSymbol{\umu}     {0}{upmath}{16}
      \NewMathSymbol{\upartial}{0}{upmath}{40}
      \NewMathSymbol{\leqslant}{3}{AMSa}{36}
      \NewMathSymbol{\geqslant}{3}{AMSa}{3E}

      \let\leq=\leqslant 
      \let\geq=\geqslant 
    \fi
  \fi
\fi 

\ifnfssone
  \newmathalphabet{\mathit}
  \addtoversion{normal}{\mathit}{cmr}{m}{it}
  \addtoversion{bold}{\mathit}{cmr}{bx}{it}
  \newmathalphabet{\mathbfit} 
  \addtoversion{normal}{\mathbfit}{cmr}{bx}{it}
  \addtoversion{bold}{\mathbfit}{cmr}{bx}{it}
  \newmathalphabet{\mathbfss} 
  \addtoversion{normal}{\mathbfss}{cmss}{bx}{n}
  \addtoversion{bold}{\mathbfss}{cmss}{bx}{n}
  \ifAMStwofonts
    \ifCUPmtlplainloaded \else
      \UseAMStwoboldmath
      \makeatletter
      \new@mathgroup\upmath@group
      \define@mathgroup\mv@normal\upmath@group{eur}{m}{n}
      \define@mathgroup\mv@bold\upmath@group{eur}{b}{n}
      \edef\UPM{\hexnumber\upmath@group}
      \new@mathgroup\amsa@group
      \define@mathgroup\mv@normal\amsa@group{msa}{m}{n}
      \define@mathgroup\mv@bold\amsa@group{msa}{m}{n}
      \edef\AMSa{\hexnumber\amsa@group}
      \makeatother
      \mathchardef\upi="0\UPM19
      \mathchardef\umu="0\UPM16
      \mathchardef\upartial="0\UPM40
      \mathchardef\leqslant="3\AMSa36
      \mathchardef\geqslant="3\AMSa3E

      \let\leq=\leqslant 
      \let\geq=\geqslant 
    \fi
  \fi
\fi 

\ifnfsstwo
  \DeclareMathAlphabet{\mathbfit}{OT1}{cmr}{bx}{it}
  \SetMathAlphabet\mathbfit{bold}{OT1}{cmr}{bx}{it}
  \DeclareMathAlphabet{\mathbfss}{OT1}{cmss}{bx}{n}
  \SetMathAlphabet\mathbfss{bold}{OT1}{cmss}{bx}{n}
  \ifAMStwofonts
    \ifCUPmtlplainloaded \else
      \DeclareSymbolFont{UPM}{U}{eur}{m}{n}
      \SetSymbolFont{UPM}{bold}{U}{eur}{b}{n}
      \DeclareSymbolFont{AMSa}{U}{msa}{m}{n}
      \DeclareMathSymbol{\upi}{0}{UPM}{"19}
      \DeclareMathSymbol{\umu}{0}{UPM}{"16}
      \DeclareMathSymbol{\upartial}{0}{UPM}{"40}
      \DeclareMathSymbol{\leqslant}{3}{AMSa}{"36}
      \DeclareMathSymbol{\geqslant}{3}{AMSa}{"3E}

      \let\leq=\leqslant 
      \let\geq=\geqslant 
    \fi
  \fi
\fi 

\ifCUPmtlplainloaded \else
  \ifAMStwofonts \else 
    \def\upi{\pi}
    \def\umu{\mu}
    \def\upartial{\partial}
  \fi
\fi

\begin{document}
\title{The Isotopic Mixture of Barium in the Metal-poor Subgiant HD 140283}
\author[David L. Lambert and Carlos Allende Prieto]
       {David L. Lambert and Carlos Allende Prieto\\ 
       Department of Astronomy, University of Texas, Austin, TX 78712-1083, USA\\}
\date{Accepted .
      Received ;
      in original form 2002 }

\pagerange{\pageref{firstpage}--\pageref{lastpage}}
\pubyear{2001}

\maketitle

\label{firstpage}

\begin{abstract}

Analyses of the abundances of neutron-capture elements
 have led to the belief that these elements
in metal-poor stars are  $r$-process
products with  relative abundances  closely resembling 
those found in the solar system.
This picture  was challenged by Magain (1995), who found that a
pure  $r$-process mix of the barium isotopes was inconsistent
with the mix of odd to even barium isotopes derived from
analysis of the Ba {\sc ii} line at 4554 \AA\ in the spectrum of
the metal-poor subgiant HD 140283.
In this paper, we  address Magain's challenge
using new high resolution high signal-to-noise
spectra of HD 140283, and find, in contrast to his result,
that a solar-like $r$-process
isotopic mixture  provides a fair fit to the observed 4554 \AA\ profile.

\end{abstract}


\section{Introduction}

Nucleosynthesis of heavy elements is dominated by the two neutron-capture
processes known as the $s$- and the $r$-process.
Some nuclides can be synthesized by
only the $s$-process, and others by only  
the $r$-process. Both processes
contribute  almost all the  remaining nuclides, with the exception of
a few heavy neutron-poor nuclides
of low abundance termed $p$-nuclei, which 
cannot be synthesized by either process.

 The He-shell of a
star in the asymptotic giant branch (AGB) 
has been identified as the site responsible for the
{\it main} $s$-process, which
synthesizes nuclides heavier than about strontium.
Of the observable elements, barium and heavier elements are
to differing degrees produced through the {\it main} $s$-process.
What is known as the {\it weak} $s$-process synthesizes nuclides
between the iron-group elements and about strontium. Operation
of the {\it weak} $s$-process is considered to occur in the helium and
carbon shells of massive stars in their hydrostatic burning phases.
The $r$-process is likely associated with the deep
interior of  Type II
supernovae.

Since massive stars dying as Type II supernovae evolve faster than
the stars that populate and evolve from the
 AGB, one anticipates that
very metal-poor stars,
formed from the interstellar medium (ISM)
 when it was little polluted with stellar
ejecta, will contain $r$-process
products but little to no  {\it main} $s$-process products.
Products of the {\it weak} $s$-process may accompany $r$-process
products but their abundance is likely very low. As the Galaxy aged and
the ISM became more polluted, the stars that formed would have  contained
more metals and
both $r$- and $s$-process products.
In this picture, one expects the heavy elements in the most metal-poor stars to
be exclusively  $r$-process products, a suggestion made by
Truran (1981).

Magain (1995, see also Magain \& Zhao 1993a)
claimed, on the basis of published elemental abundances,
that ``there is no secure
observational evidence in support of Truran's suggestion''. He
applied a novel test to the metal-poor
subgiant HD 140283 to determine the relative
contributions of the $r$- and $s$-process, and used the result
to reinforce his claim. His test, which had been anticipated
by Cowley \& Frey (1989), involved an analysis of
the Ba\,{\sc ii} 4554 \AA\ line profile 
to determine 
the fractional abundance of the odd isotopes.
Contributions of the  various isotopes cannot be distinguished in
the  4554 \AA\ line's profile solely on the basis of the very small
isotopic wavelength shifts, but the hyperfine splittings
 (about -30 m\AA\ to + 20m\AA) contributed by
 the two odd isotopes
result in a broadening of the stellar line that is dependent on the
abundance ratio of the odd to even isotopes.
 According to the
analyses of the solar system abundances, the $s$- and $r$-process
contribute quite differently to the  fractional abundance of the odd
isotopes, that is $f_{\rm odd} = [N(^{135}$Ba) + $N(^{137}$Ba)]/$N$(Ba)
where
$f^s_{\rm odd}$ = 0.11 but $f^r_{\rm odd}$ = 0.46, according
to Arlandini et al. (1999). Thus, the line width,
which offers a way to measure $f_{\rm odd}$,  is a potential
indicator of the relative contributions of the $s$- and $r$-process.

Magain obtained and analysed a high-resolution (R $\equiv
\lambda/\Delta\lambda$
 = 100,000) high signal-to-noise (S/N $\simeq$ 400) observation 
 of the 4554 \AA\ line in
the spectrum of  HD\,140283. Using the Fe\,{\sc ii}
line at 4555.9 \AA\ to determine the broadening of a line 
 unaffected by
isotopic and hyperfine splitting, Magain's
fit to the 4554 \AA\ line gave the fractional abundance of the
odd Ba isotopes as $f_{\rm odd} = 0.06 \pm 0.06$.
On the basis of the solar estimates of
$f^s_{\rm odd}$ and $f^r_{\rm odd}$, 
Magain concluded that HD 140283's barium
isotopic ratio reflects `pure $s$-process production' and `significant
enhancement of the $r$-process contribution' is excluded.
This conclusion reinforced his critique of
published abundance analyses of metal-poor stars.

In this paper, we reobserve and
 analyse afresh HD 140283's  Ba\,{\sc ii} 4554 \AA\ line profile.
 Our new spectra,  superior to the
material previously published,
are described in Section 2. The isotopic analysis
 is discussed in Section 3, and the paper
concludes with a discussion of the barium isotopic mix
in the context of the heavy element composition of HD 140283 and
other metal-poor stars.

\section{Observations}

All of our observations of HD 140283  were carried out with
the Harlan J. Smith Telescope  and  the {\it 2dcoud\'e}
cross-dispersed echelle spectrograph
(Tull et al. 1995) at the W.J. McDonald Observatory
(Mt. Locke, West Texas). Spectra with a FWHM resolving power
$R  \simeq 200,000$ were secured
on  May  23-25, 1997.  A series of
1/2 hour exposures were  cross-correlated and combined
to reach a signal-to-noise ratio  $\sim 550$ per pixel at the
Ba\,{\sc ii} line.
The detector was  a 2k$\times$2k Tektronix CCD with a very large
dynamic range.
We refer the reader to Allende Prieto (1998) and
Allende Prieto et al. (1999) for more information about these data.

Observations  at
a resolving power $R  \simeq 60000$
and with a S/N
 around 200 were acquired in 1993 and 1994
by Ram\'on J. Garc\'{\i}a L\'opez with a 1k$\times$1k Tektronix
CCD.
 These
spectra provided equivalent widths for 64 iron lines that
were used to constrain the microturbulence.
Equivalent
 widths were measured  by Gaussian fits to individual
 or slightly blended lines.

\section{Line Profile Analysis}

The profile of the Ba\,{\sc ii} 4554 \AA\ line is set by  isotopic and hyperfine
splittings, and by physical conditions in the stellar
atmosphere such as turbulent motions and the stellar rotation.
We use model atmospheres and our high-resolution
spectra to constrain the stellar contributions to the line
profile.

 The isotopic and hyperfine splittings have been
measured by laboratory spectroscopists.
Rutten
(1978) collated the  measurements.
 McWilliam (1998)
used these data to calculate the  components of the 4554 \AA\ line
(and other  lines). He ignored the isotopic
wavelength shifts of the even isotopes,
 but this is acceptable here: the $^{134}$Ba - $^{138}$Ba shift is only
about 2.4 m\AA.
Then, even isotopes produce the
single component that we have
centred, following McWilliam, at 4554.000 \AA.
No rational distinction
can be made between the two odd isotopes.
 The true central
wavelength is not important, given the uncertainties in the radial velocity
of the star, convective motions, oscillations, etc., and we allow for
a velocity shift between the observed and calculated profiles.
The relative intensities of the hyperfine splitting (hfs) 
components of the 4554 \AA\ line
published by McWilliam are
 in good agreement with the calculation of Cowley \& Frey (1989)
 and our own.
 (We  adopt McWilliam's choice for the
line's $gf$-value [$\log gf = +0.17$].)

Our  analysis used a classical  model atmosphere constructed for the
standard assumptions of plane parallel homogeneous layers in hydrostatic,
flux, and local thermodynamic equilibrium.
The  model was 
taken from the Kurucz (1992)
  grid of non-overshooting models. By interpolation, we obtained
a model atmosphere for the parameters:
$T_{\rm eff} = 5777$ K, $\log g = 3.74$, and [Fe/H]$=-2.7$, which are
very similar to the values recommended by Snider et al. (2001) 
based on the Infrared flux method and the parallax measured by 
{\it Hipparcos}.
 In Table 1, we list the adopted parameters and the abundances found in 
 some of the previous analyses. 
Our metallicity in Table 1 corresponds to the iron abundance we derive below,
scaled to the solar value  we find from a
parallel analysis of the solar flux spectrum of Kurucz et al. (1984).
 As pointed out by Magain (1995), the
 predicted Ba\,{\sc ii} line profiles
are relatively 
insensitive to the particular choice of stellar parameters.
 We  calculated  synthetic spectra using the LTE code
  MOOG (Sneden 1973), and checked the results with
a second LTE code  MISS (Allende Prieto et al. 1998), noticing negligible
differences between their outputs.

\begin{table}
\scriptsize
\centering
\begin{minipage}{180mm}
\caption{HD\,140283: Results of Spectroscopic Analyses}
\begin{tabular}{crcrrrrr} \hline
Authors &$T_{\rm eff}$ &$\log
g $&[Fe/H] &[Sr/Fe] &[Y/Fe]
&[Ba/Fe] &[Eu/Fe] \\
&(K) &(cgs)& & & & & \\
 \hline
{MMZ}\footnote{Magain (1989) and Magain \& Zhao (1990): also [La/Fe] = $-$0.05} & 5640 & 3.10 & $-$2.8 & $-$0.34 & $-$0.58 & $-$1.13 & $+$0.21\\
{GS}\footnote{Gratton \& Sneden (1994)} & 5690 & 3.58 & $-$2.6 & $-$0.03 & $-$0.38 & $-$0.64 & $+$0.09\\
{RNB}\footnote{Ryan, Norris, \& Beers (1996)} & 5750 & 3.40 & $-$2.5 & $-$0.44 & $+$0.18 & $-$0.91 & ...\\
{MGB}\footnote{Mashonkina, Gehren, \& Bikmaev (1999)} & 5640 & 3.65 & $-$2.3 & ... & ... & $-$0.80 & ...\\
{F}\footnote{Fulbright (2000)} & 5650 & 3.40 & $-$2.4 & ... & ... & $-$1.03 & ...\\
{MK}\footnote{Mishenina \& Kovtyukh (2001)} & 5650 & 3.5 & $-$2.4 & ... & $-$0.19 & $-$0.78 & ...\\
{LAP}\footnote{This paper} & 5777 & 3.74 & $-$2.4 & ... & ... & $-$1.09 & ...\\
\hline
\end{tabular}
\end{minipage}
\end{table}

\begin{figure*}
\centering
\includegraphics[width=12cm,angle=0]{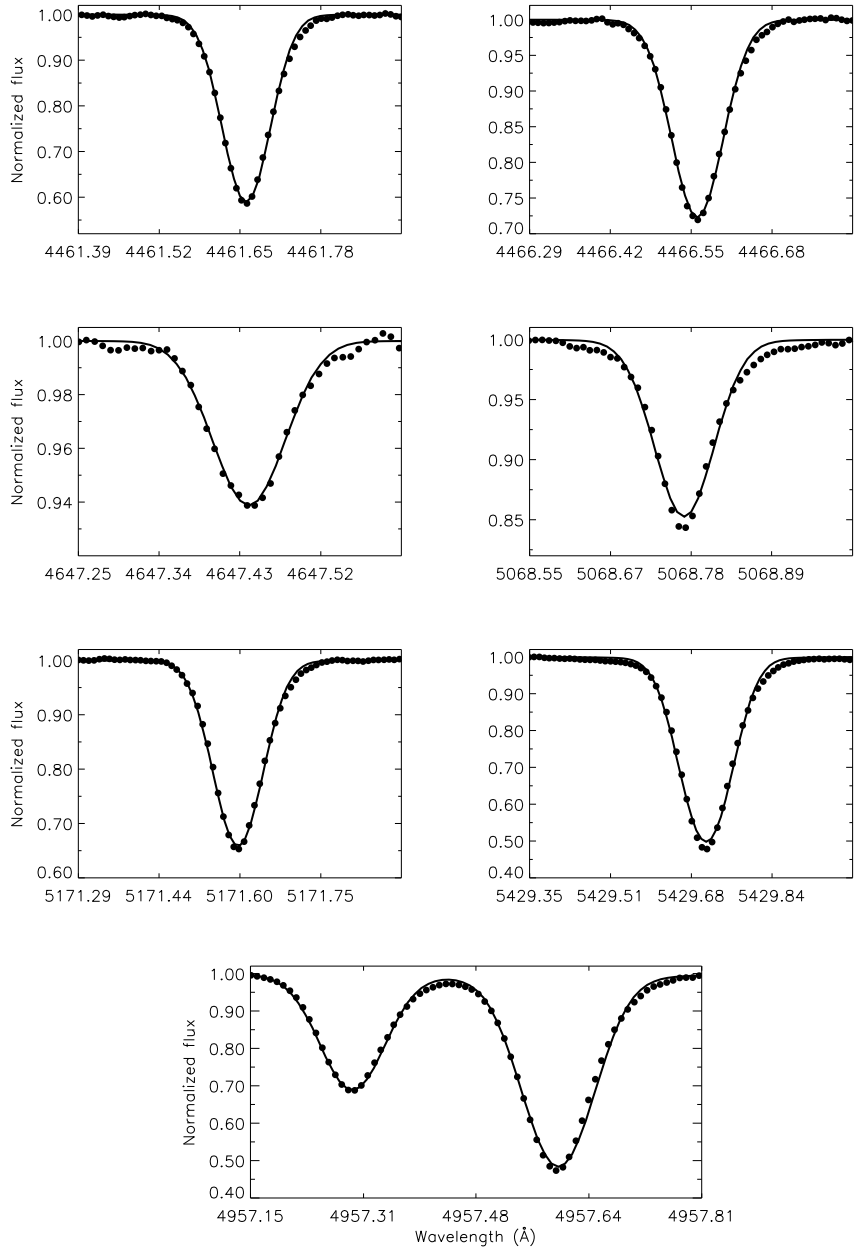}
\protect\caption[ ]{
Best-fit synthetic profiles (thick solid) compared to the observed
spectra for eight iron lines used to determined the velocity and
instrumental line broadening in the spectrum of HD 140283. 
\label{hdfwhm}}
\end{figure*}

With the chosen model atmosphere,
we  analysed the equivalent widths of iron lines
 measured off the $R=60,000$ spectra, finding that a microturbulence of
 $\xi=1.42$ km s$^{-1}$ made the iron abundance 
 independent from the line equivalent width to yield 
$\log \epsilon ({\rm Fe}) = 5.18 \pm$  0.11
from 51 Fe\,{\sc i}
 lines
($gf$-values from Oxford; e.g., Blackwell et al. 1986)
and $\log \epsilon ({\rm Fe}) = 5.12 \pm  0.08$ ($gf$-values from
Giridhar \& Arellano Ferro 1994)
from 13 Fe\,{\sc ii} lines. 
 Here and subsequently, we have
adopted the standard deviation ($\sigma$) as the 
uncertainty.
 The equivalent widths, atomic data, and abundances for the individual lines
  appear in Table 2. Note that the 
  ionisation equilibrium of iron is rather
well satisfied by this analysis.

Adopting the above microturbulence, we  analysed a
sample
of Fe\,{\sc i} lines observed at $R \simeq 200,000$ (see Table 3).
 The transition
probabilities of these lines have been measured by O'Brian et al. (1991)
and/or at Oxford.
We achieved the best fit to the line profiles by varying the iron abundance,
the central wavelength
of the line, and the FWHM of an assumed-Gaussian broadening. The Gaussian
is intended to account for the rotational, macroturbulent,  and
instrumental broadening.
Collisional broadening by hydrogen atoms is considered using
the theory and
parameters prescribed by Barklem, Piskunov, \& O'Mara (1999), and references
therein. Approximate formulae were used for the less important Stark
and natural broadening.
Table 3 lists  the best-fitting parameters, while
Fig. 1 illustrates the fitting graphically.
This set of Fe\,{\sc i} lines give
 $\log \epsilon ({\rm Fe}) = 5.09 \pm 0.09$,
and a FWHM of 5.06  $\pm 0.33$ km s$^{-1}$  for the Gaussian
broadening.  A fit to the Fe\,{\sc ii} line used by
Magain gives FWHM = 5.5 km s$^{-1}$, a value within the scatter
of results from the Fe\,{\sc i} lines.
The fits were accomplished by the  Nelder-Mead simplex method (Nelder \& Mead
1965), as implemented by Press et al. (1988).
It is interesting to note that we obtain a consistent iron abundance from the
analysis of the equivalent widths in Table 3, which were measured by fitting
a Gaussian to the observed profiles. Replacing the damping 
parameters  calculated by Barklem et al. (1999) by the Uns\"old approximation, 
our abundance estimate  would increase by 0.04 dex.

\begin{table}{h}
\centering
\scriptsize
\caption{Spectral Lines Selected to determine the  
microturbulence  in HD 140283}
\begin{tabular}{ccccc}
\hline
Wavelength  & EP & $\log gf$ &  W$_{\lambda}$ & $\log \epsilon({\rm Fe})$ \\
 (\AA)  & (eV) &   (dex)   &  (m\AA) &   (dex) \\ 
\hline
\multicolumn{5}{c}{Fe I} \\
\hline
3786.68  &  1.01  &  $-$2.23  &  40  &  5.18 \\
3790.10  &  0.99  &  $-$1.76  &  54  &  5.05 \\
3833.31  &  2.56  &  $-$1.03  &  20  &  5.02 \\
3906.48  &  0.11  &  $-$2.24  &  79  &  5.42 \\
3917.18  &  0.99  &  $-$2.16  &  44  &  5.17 \\
4005.25  &  1.56  &  $-$0.61  &  79  &  5.24 \\
4067.27  &  2.56  &  $-$1.42  &  11  &  5.07 \\
4147.67  &  1.49  &  $-$2.10  &  24  &  5.11 \\
4152.17  &  0.96  &  $-$3.23  &  9  &  5.18 \\
4172.75  &  0.96  &  $-$3.07  &  16  &  5.29 \\
4174.92  &  0.92  &  $-$2.97  &  16  &  5.15 \\
4177.60  &  0.92  &  $-$3.06  &  24  &  5.50 \\
4202.03  &  1.49  &  $-$0.71  &  78  &  5.18 \\
4216.19  &  0.00  &  $-$3.36  &  39  &  5.21 \\
4222.22  &  2.45  &  $-$0.97  &  31  &  5.08 \\
4250.12  &  2.47  &  $-$0.41  &  52  &  5.06 \\
4271.16  &  2.45  &  $-$0.35  &  60  &  5.17 \\
4337.05  &  1.56  &  $-$1.70  &  35  &  5.03 \\
4375.93  &  0.00  &  $-$3.03  &  55  &  5.24 \\
4442.34  &  2.20  &  $-$1.26  &  30  &  5.10 \\
4447.72  &  2.22  &  $-$1.34  &  25  &  5.07 \\
4489.74  &  0.12  &  $-$3.97  &  11  &  5.15 \\
4494.57  &  2.20  &  $-$1.14  &  34  &  5.08 \\
4733.60  &  1.49  &  $-$2.99  &  6  &  5.18 \\
4939.69  &  0.86  &  $-$3.34  &  10  &  5.16 \\
4994.13  &  0.92  &  $-$3.08  &  14  &  5.13 \\
5012.07  &  0.86  &  $-$2.64  &  32  &  5.13 \\
5079.23  &  2.20  &  $-$2.07  &  9  &  5.19 \\
5079.74  &  0.99  &  $-$3.22  &  13  &  5.29 \\
5083.34  &  0.96  &  $-$2.96  &  16  &  5.10 \\
5107.45  &  0.99  &  $-$3.09  &  12  &  5.13 \\
5107.65  &  1.56  &  $-$2.42  &  14  &  5.10 \\
5110.41  &  0.00  &  $-$3.76  &  25  &  5.20 \\
5123.72  &  1.01  &  $-$3.07  &  12  &  5.14 \\
5151.91  &  1.01  &  $-$3.32  &  7  &  5.11 \\
5194.94  &  1.56  &  $-$2.09  &  26  &  5.13 \\
5198.71  &  2.22  &  $-$2.14  &  9  &  5.24 \\
5307.36  &  1.61  &  $-$2.99  &  6  &  5.31 \\
5397.13  &  0.92  &  $-$1.99  &  64  &  5.26 \\
5405.78  &  0.99  &  $-$1.84  &  67  &  5.24 \\
5429.70  &  0.96  &  $-$1.88  &  68  &  5.29 \\
5434.53  &  1.01  &  $-$2.12  &  53  &  5.18 \\
5701.55  &  2.56  &  $-$2.22  &  3  &  5.06 \\
6065.49  &  2.61  &  $-$1.53  &  10  &  5.04 \\
6136.62  &  2.45  &  $-$1.40  &  20  &  5.11 \\
6137.70  &  2.59  &  $-$1.40  &  17  &  5.16 \\
6219.29  &  2.20  &  $-$2.43  &  3  &  4.99 \\
6252.56  &  2.40  &  $-$1.69  &  14  &  5.17 \\
6265.14  &  2.18  &  $-$2.55  &  5  &  5.29 \\
6593.88  &  2.43  &  $-$2.42  &  6  &  5.44 \\
6750.15  &  2.42  &  $-$2.62  &  3  &  5.36 \\
%
\hline
\multicolumn{5}{c}{Fe II} \\
\hline
4173.49  &  2.58  &  $-$2.18  &  27  &  5.02 \\
4178.86  &  2.58  &  $-$2.48  &  20  &  5.14 \\
4489.19  &  2.82  &  $-$2.96  &  5  &  5.13 \\
4491.41  &  2.85  &  $-$2.70  &  8  &  5.10 \\
4555.90  &  2.83  &  $-$2.29  &  17  &  5.06 \\
4303.17  &  2.70  &  $-$2.49  &  19  &  5.22 \\
4416.83  &  2.77  &  $-$2.55  &  11  &  5.05 \\
4583.85  &  2.80  &  $-$1.84  &  42  &  5.19 \\
4731.46  &  2.89  &  $-$2.92  &  4  &  5.03 \\
4923.93  &  2.89  &  $-$1.24  &  57  &  5.01 \\
5234.64  &  3.22  &  $-$2.24  &  13  &  5.22 \\
5316.62  &  3.15  &  $-$1.85  &  24  &  5.10 \\
6456.40  &  3.90  &  $-$2.20  &  4  &  5.22 \\
\hline
\end{tabular}
\end{table}

Next, we fit the profile of the Ba\,{\sc ii} 4554 \AA\ line 
by changing the barium abundance, and 
 $f_{\rm odd}$.
  Our ultra-high
resolution spectra  can easily distinguish the 
effects of the two parameters;
at the metallicity of HD 140283, the equivalent width is almost
independent of $f_{\rm odd}$.
We found that a linear relationship of the form 
$f_{\rm odd} = 0.51 - 0.51 ({\rm FWHM} - 4.65)$ represents very well the
dependence of the derived $f_{\rm odd}$ on the magnitude of the Gaussian 
broadening. For FWHM $= 5.06 \pm 0.33$ km s$^{-1}$, we find our best fit
(reduced $\chi^2$ = 2.9)
 with $\log \epsilon ({\rm Ba}) = -1.18 \pm 0.02$ 
dex and $f_{\rm odd}= 0.30 \pm 0.17$. 
Note that our FWHM for the Fe\,{\sc ii} line used by Magain would give
a value of $f_{\rm odd}$ ($\simeq$ 0.08) close to his value.
The FWHM of the Ba\,{\sc ii} line appears not to be affected by blending lines.
Cowley \& Frey searched for  potential 
blends that might be disturbing the 
profile in a solar-like star, concluding that only two transitions 
Cr\,{\sc i} 4553.94 \AA\ and 
Zr\,{\sc ii} 4553.97 \AA\ were likely to affect the barium resonance line.
Our test calculation using the atomic parameters in Kurucz's library of lines
shows that these two features are irrelevant in the case of HD 140283;
even if the $gf$-values are increased by an order of magnitude the pair
 together cannot contribute more than
0.1 m\AA,  or about 0.5 \% of the
equivalent width of the barium line.
Our Ba abundance is derived assuming LTE but non-LTE calculations
by Mashonkina, Gehren, \& Bikmaev (1999) show that the non-LTE abundance
will not be very different.

The best-fit to the observed
profile is shown in Fig. 2. The differences between the observed and
fitted profile are symmetric about the line's centre. 
Using Magain's hfs model, the result for $f_{\rm odd}$ does not change 
significantly. 
Apart from the uncertainty in FWHM, there are other contributors
to the error budget for $f_{\rm odd}$. As the star is very metal-poor and the
quality of the spectrum is very high, we deem the continuum determination to be
 reliable. If we adopt the $1\sigma$ uncertainty for S/N= 550, which
is 0.002, this factor can alter the derived $f_{\rm odd}$ by no more than 0.05.
Changes in $T_{\rm eff}$ and [Fe/H] 
in the adopted model atmosphere can
shift the barium abundance significantly, but
they have no effect on $f_{\rm odd}$. 
The adopted microturbulence is relevant:  a change of
 $\pm 0.2$ km s$^{-1}$ changes $f_{\rm odd}$  by  $\mp 0.11$.
The adopted gravity enters into the determination of $f_{\rm odd}$  through
the pressure broadening of the line, but
a variation of 0.2 dex in $\log g$, which can be taken as
a conservative estimate for HD 140283, shifts $f_{\rm odd}$ by only 0.01.
This error source  should be further reduced by our procedure of fitting 
FWHM with a set of lines observed simultaneously and modelled with 
the same model atmosphere.
In summary, we assign an uncertainty of about 
 $\sqrt{0.17^2 + 0.05^2 + 0.11^2 + 0.01^2} \simeq 0.21$ to our
 determination of $f_{\rm odd}$.

\begin{table*}
\centering
\caption{Fe {\sc i} spectral lines selected to determine the 
line broadening  in HD 140283. The Fe\,{\sc ii} 4555.9 \AA\
line used by Magain is listed last. $\sigma_0$ is the cross-section for
broadening by collisions with neutral hydrogen for a velocity of the perturber
$v_0=10^4$ m s$^{-1}$, and  $\sigma(v) \propto v^{\alpha}$ 
(see, e.g., Barklem \& O'Mara 1997).}
\begin{tabular}{ccccccccc}
\hline
Wavelength & EP & $\log gf$ & $\sigma_0$ & $\alpha$ 
 & FWHM & $\log \epsilon({\rm Fe})$ & W$_{\lambda}$ &reduced $\chi^2$ \\
 (\AA)  & (eV) &   (dex)  & (a$_0^2$) &  &  (km s$^{-1}$)  &   (dex) & (m\AA) & \\ 
\hline
4461.66 &0.087 & $-$3.19 & 217 & 0.250 & 4.7   &5.12 & 41 & 10.5   \\
4466.56 &2.831 & $-$0.60 & 222 & 0.263 & 5.1   &5.03 & 29 & 5.7   \\
4647.44 &2.949 & $-$1.35 & 283 & 0.259 & 5.1   &5.06 & 6 & 0.9   \\
4957.30 &2.851 & $-$0.41 & 727 & 0.238 & 5.5   &4.98 & 37 & 16.3  \\
4957.60 &2.808 & $+$0.23 & 714 & 0.238 & 5.5   &4.98 & 70 & 16.3  \\
5068.77 &2.940 & $-$1.04 & 738 & 0.237 & 4.6   &5.18 & 16 & 7.1   \\
5171.61 &1.485 & $-$1.79 & 281 & 0.253 & 5.1   &5.08 & 41 & 3.3  \\
5429.71 &0.958 & $-$1.88 & 240 & 0.248 & 4.9   &5.24 & 68 & 18.4  \\
\hline
4555.89 &2.828 & $-2.29$ & ... & ...   & 5.5   & 5.00 & 16 & 9.8 \\
\hline
\end{tabular}
\end{table*}

Several potential sources of error have been considered, but there are
more caveats.
Velocity fields might introduce differences between 
the FWHM of the comparison Fe\,{\sc i} lines and the Ba\,{\sc ii} line.
The set of Fe\,{\sc i} lines  spans a range of equivalent widths;
some are weaker but most are slightly stronger than the barium line (20 m\AA). 
 Our data, however, does not show a correlation of FWHM with line 
strength. 
Interestingly, Ryan et al. (2002) have recently reported correlations between
 line width and  equivalent width for Fe\,{\sc i} lines
 in metal-poor stellar spectra, in the sense that stronger lines are wider.  
The correlation was  demonstrated for lines stronger than 30 m\AA. 
In the hypothetical case that such a pattern were present in HD 140283's lines,
and that the barium line followed the same trend as the iron lines, 
the fact that most lines in Table 3 are stronger than the Ba\,{\sc ii} 
feature could induce a bias in the derived $f_{\rm odd}$. Accounting for
such an effect would imply a higher $f_{\rm odd}$ than we derive,  in
closer agreement with $f^r_{\rm odd}$.

Close examination of the profiles of unblended lines 
 shows that they are asymmetric, and strong lines are
redshifted relative to weak lines (Allende Prieto et al. 1999).
Theoretical profiles from a classical atmosphere are necessarily
symmetric and unshifted relative to one another (pressure
shifts are too small to be detected). It is the convective motions
associated with stellar granulation  that induce the line
asymmetries and shifts. Inspection of Figure 1 shows that Fe\,{\sc i}
lines are asymmetric with stronger red than blue wings. 
A line like the Ba\,{\sc ii} line, if
free from isotopic and hyperfine splittings, would also have
wings red-shifted relative to the line core. The observed line is,
as Figure 2 shows, quite symmetric. We may attribute this
symmetry to the combination of the asymmetry
due to granulation in  the presence of the odd isotopes and the asymmetry of their
hyperfine splittings about line centre.  It would, nonetheless, be
useful to repeat the analyses of the lines using 3D hydrodynamical
simulations of stellar granulation.

Another source of possible error might be linked to our use of
LTE line profiles.
Non-LTE calculations for Ba\,{\sc ii} line formation in classical
atmospheres were presented by Mashonkina \& Gehren (2000). For stars
as metal-poor as HD\,140283, the non-LTE Ba abundance from the 4554 \AA\ line
 is slightly larger than
the LTE value, i.e., the non-LTE line for a given abundance is
weaker than the equivalent LTE line.  In the case of HD\,140283, 
using a model with T$_{\rm eff}$ = 5640 K, $\log g$ = 3.65,
and [Fe/H] $= -2.30$,
Mashonkina \& Gehren find the non-LTE abundance from 4554 \AA\ to be 0.15 dex
larger than the LTE abundance. Our choices for atmospheric
parameters are different and will affect slightly the non-LTE correction,
but it is impossible to estimate the magnitude of the change from
the limited results given  by Mashonkina \& Gehren.

\begin{figure*}
\centering
\includegraphics[width=12cm,angle=0]{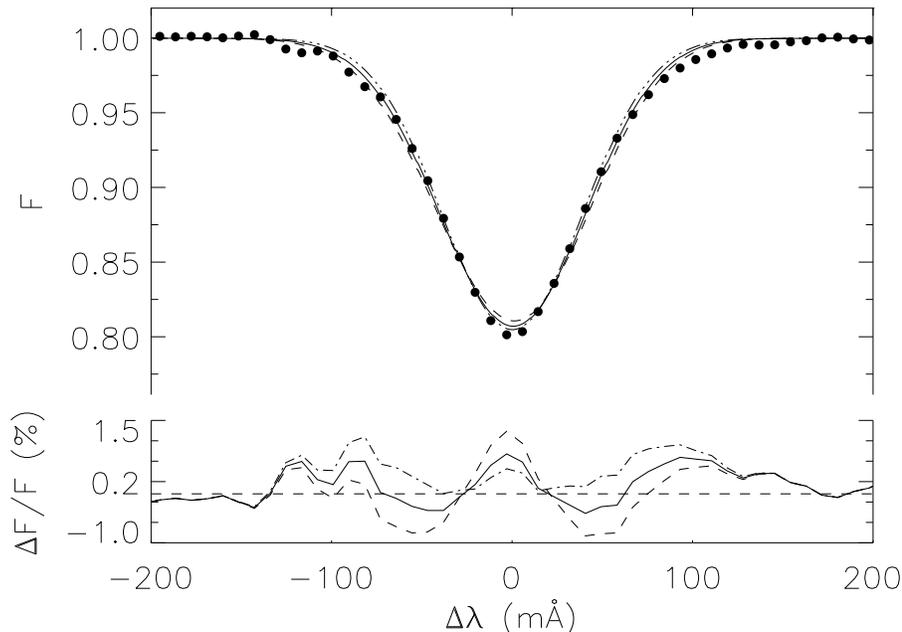}  
\protect\caption[ ]{
Best-fit ($f_{\rm odd}=0.31$) synthetic profile (thick solid line) and the observed 
profile of the
Ba\,{\sc ii} 4554.0 \AA\ line in HD 140283. The dashed and dot-dashed
 lines correspond to a change of $\pm 0.21$ in $f_{\rm odd}$.
\label{hdba}}
\end{figure*}

The small change in Ba abundance is not a critical factor. It is the changes
in the line profile  that may affect the determination of
$f_{\rm odd}$.
The NLTE 
source function for the 4554 \AA\ line is predicted to
be greater than the local Planck function over the
region of line formation,  the number density of
ground state ions is reduced below the LTE value,  and  the
line is weakened slightly. 
One may suppose that the weakening is
more severe at the line centre than in the wings, that is the
non-LTE line assumes a larger width for a given central depth
than for the comparable LTE line. If the observed line is then
analysed with non-LTE line profiles, the
derived $f_{\rm odd}$ would be slightly smaller than that from the LTE
calculations.

Effects of non-LTE on the Fe\,{\sc i} comparison
 lines should also be assessed but
unfortunately published predictions seem too incomplete to
provide a quantitative assessment. Th\'{e}venin \& Idiart (1999), who performed
non-LTE calculations for iron, found the Fe abundance of HD 140283 to increase
by 0.29 dex when non-LTE replaced LTE in the analysis of Fe\,{\sc i}
and Fe\,{\sc ii} lines, and the determined surface gravity was in good
agreement with the {\it Hipparcos}-based value. The leading non-LTE effect
is an overionisation of neutral iron.
 It is impossible from the
brief published description of the calculations for HD 140283 to determine the
outcome on the line profiles, but it seems likely that the non-LTE
profiles are slightly
shallower than their LTE equivalents, as was the case for the
Ba\,{\sc ii} 4554 \AA\ line. Since we have used six different lines,
we suppose that the the net effect is too small
to affect the mean FWHM, although it could definitely be a major
source of scatter. In addition, it is clear that
Th\'{e}venin \& Idiart's calculations are not the last word on the
possible departures
from LTE, which may be smaller than calculated by
them (Fulbright 2000; Gehren et al. 2001). Indeed, our LTE analysis
using the {\it Hipparcos}-based gravity gives consistent
abundances from the Fe\,{\sc i} and Fe\,{\sc ii} lines.
 
 It is of interest to perform a similar analysis of the Ba {\sc ii} 
 line at 4554 \AA\ in the solar spectrum. In the spectrum of a star like 
 the Sun, this line
 is very strong. Its core, formed in high atmospheric layers and expected
 to suffer serious departures from LTE, is excluded from the fitting process. 
 This experiment, 
 mirroring the analysis of HD 140283 and using the Fe {\sc i} lines at
 4602.0, 5247.1, 6151.6, 6200.3, and 6750.2 \AA\ to find the iron abundance
 ($\log \epsilon$ (Fe) = $7.48 \pm 0.04$ dex)
 and  FWHM ($4.15 \pm 0.12$ km s$^{-1}$), 
 yields a best-fit result of 
 $\log \epsilon$(Ba) $= 2.32$ and $f_{\rm odd} = 0.20$.
 These figures are  consistent with the meteoritic abundance
 $\log \epsilon$(Ba) $= 2.22$ and isotopic fraction 
 $f_{\rm odd} = 0.18$. In fact, the close agreement between 
 the derived and meteoritic $f_{\rm odd}$, and the larger discrepancy for the
 Ba abundance, 
 seem to confirm our expectation of a much higher accuracy in the
 determination of the first quantity.

In summary, our analysis of HD 140283's
 Ba\,{\sc ii} 4554 \AA\ profile gives $f_{\rm odd} =  0.30 \pm 0.21$,
a result consistent with the solar $r$-process contributions to barium
which give $f^r_{\rm odd} = 0.46$. Our analysis seems not to confirm
 Magain's
(1995) conclusion 
 that the odd Ba isotopes have a low value ($f_{\rm odd} = 0.06 \pm 0.06$)
in the atmosphere of this star.
 It is clear from Figure 2 that the Ba\,{\sc ii} line profile
is  only weakly sensitive
 to $f_{\rm odd}$. That is reflected in our large error bar for this
 parameter ($\sim 70$\%). High-resolution 
 high-quality spectra
must be paired with a realistic representation of the stellar
atmosphere and the line formation to reduce the uncertainties in the isotopic
fraction provided by this method.

\section{Nucleosynthesis of Heavy Elements}

\subsection{Origins of the Solar Barium Isotopes}

Accurate isotopic abundances are available for the solar system from
laboratory measurements on meteorites.
Dissection of  the heavy element abundances into $s$- and $r$-process
contributions has been attempted many times with similar results.
 Here, we adopt
Arlandini et al.'s (1999) results from their fit of
predictions for the $s$-process operating in   low mass AGB stars.
We are interested in the relative abundance of the odd barium
isotopes and the resolution of this quantity into $s$- and $r$-process
contributions. The measured abundances give $f_{\rm odd} = 0.18$.
The $r$-process contributes only to $^{135}$Ba, $^{137}$Ba, and $^{138}$Ba,
and would by itself produce $f^r_{\rm odd}$ = 0.46.
 The $s$-process contributions would 
give $f^s_{\rm odd}$ =0.11. (The $p$-nuclei $^{130}$Ba and $^{132}$Ba
are neglected here, as is the possible but surely very small contribution
of a $p$-process to heavier Ba isotopes.)

\begin{figure*}
\centering
\includegraphics[width=8cm,angle=0]{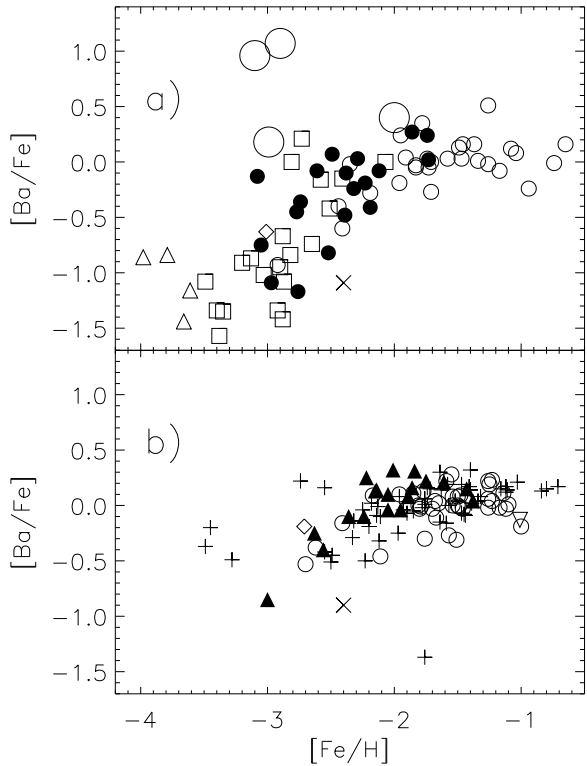}  
\protect\caption[ ]{
The abundance ratio [Ba/Fe] versus [Fe/H] for a 
representative sample of metal-poor giants (a) and 
dwarf/sub-giants (b). HD 140283 is  identified by the large
cross. The
four severely $r$-process enriched giants are shown by large
open circles. Other data points come for Figure 4a from 
Norris et al. (2001) - open triangles; McWilliam (1998) - open squares;
Johnson \& Bolte (2001) - filled circles; Fulbright (2000) - open
circles; Ryan et al. (1996) - open rhombi. For Figure 4b, the data
points are from Magain (1989) and Magain \& Zhao (1990) - filled
triangles; Fulbright (2000) - open circles; Gratton \& Sneden (1994) -
inverted open triangle; Ryan et al. (1996) - open rhombi; Stephens
\& Boesgaard (2002) - plus signs.
\label{BaFe}}
\end{figure*}

It is a remarkable empirical result that the relative
elemental abundances of the heavier $r$-process products --
barium and heavier -- appear to be almost independent of
which very metal-poor star is examined;  stars  enriched in
$s$-process products are obvious exceptions.  
This result is highlighted
by analyses of those metal-poor stars exhibiting large $r$-process
overabundances (Cowan et al.
1999; Westin et al. 2000; Hill et al. 2001; Cowan et al. 2001). 
More typical metal-poor stars also generally show a solar-like mix of the
heavier elements (Burris et al. 2000; Johnson \& Bolte 2001).
Lighter heavier elements -- Sr. Y, and Zr, for example -- show
some variation relative to heavier elements.
Adopting the assumption that  $r$-process products  for barium and
heavier elements are universal with respect to
isotopic as well as elemental abundances, we
expect $f_{\rm odd}$ $\simeq 0.5$, the solar $r$-process ratio, for
a metal-poor star in which the $r$-process dominates the
heavy element abundances.  Theoretical estimates from $r$-process
calculations suggest that $f^r_{\rm odd}$ should be accurate to
20 -- 30\% (Kratz \& Pfeiffer 2002).
 
A commonly used measure of the ratio of $s$-process to $r$-process
contributions to the composition of a metal-poor star is the Ba/Eu
ratio. 
In solar
system material, barium is
predominantly an $s$-process product, and  europium an $r$-process
product:
the $s/r$ ratio for
Ba is 81/19 but for Eu is 6/94 (Arlandini et al. 1999).
If the $r$-process relative yields of barium and europium are
solar-like, one expects [Ba/Eu] to approach $-0.7$ as the $r$-process
dominates the composition.

\subsection{HD\,140283: the [Ba/Fe]  and [Ba/Eu]  Ratios}

Although HD\,140283 has been analysed frequently, abundances have
been published for very few heavy  elements. The reason is simply that
the star is a subgiant and lines of most heavy elements are too weak
for detection. We begin by considering the Ba and Eu abundances  because
the Ba/Eu ratio is a  potent indicator of the ratio of
 $s$-process to $r$-process contributions to a chemical composition (see below).
In Table 1, we summarise  results from the
literature for the LTE abundances of Fe, Sr, Y, Ba, and Eu.
For [Ba/Fe], we obtain a mean [Ba/Fe] = $-0.9$.
The two determinations of [Eu/Fe] appear to be in good agreement: the
mean  [Eu/Fe] $= 0.15$ with [Ba/Fe] $=  -0.9$ gives [Ba/Eu] $= -1.05$. 
One might note that where the Eu abundance was measured,
the [Ba/Eu] indices are not in good agreement:
[Ba/Eu] $= -0.73$ from Gratton \& Sneden (1994)
but $-1.33$ from Magain (1989) and Magain \& Zhao (1990). This discrepancy
may suggest that a large uncertainty be attached to
the value [Ba/Eu] = $-1.05$ obtained by combining the mean [Ba/Fe] from
seven measurements with [Eu/Fe] from two measurements.

\begin{figure*}
\centering
\includegraphics[width=6cm,angle=90]{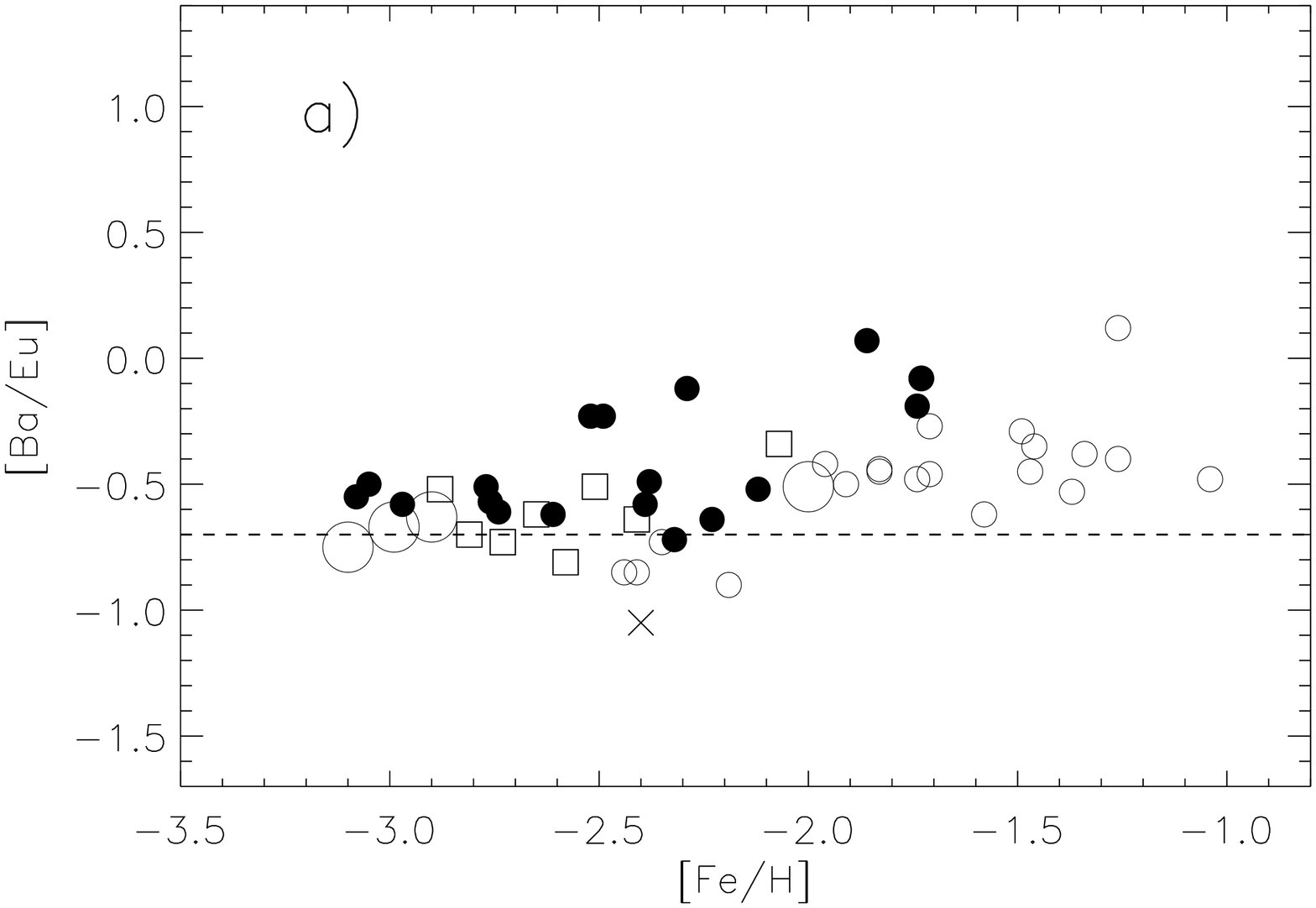}  
\includegraphics[width=6cm,angle=90]{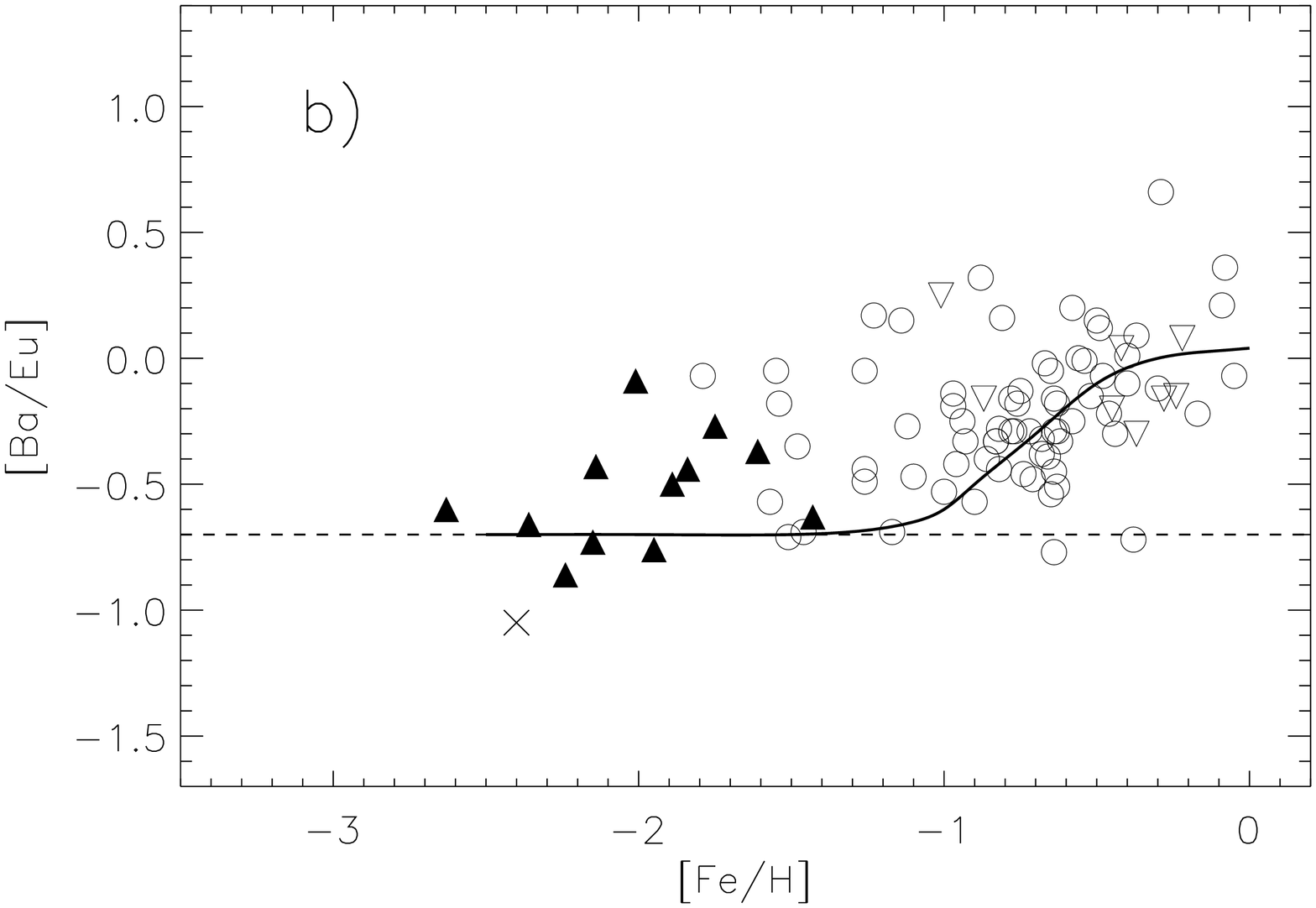}  
\protect\caption[ ]{
The abundance ratio [Ba/Eu] versus [Fe/H] for a 
representative sample of metal-poor giants (a) and 
dwarf/sub-giants (b). HD 140283 is  identified by
a  cross. Symbols are as in Figure 3 where the
four severely $r$-process enriched giants are shown by large
open circles.
\label{BaEu}}
\end{figure*}

In order to place HD 140283's  Ba and Eu abundances  
 in context,  we show the variation of [Ba/Fe]
with [Fe/H] (Figure 3) and [Ba/Eu] with [Fe/H] (Figure 4)
 for several samples of metal-poor ([Fe/H] $< -1$) stars.
Stars are separated by surface gravity in giants ($\log$ g $\leq$ 3.0)
and dwarfs and subgiants ($\log$ g $>$ 3.0).
We include in Figures 3a and 4a the  four known stars with an exceptional
$r$-process enrichment that we mentioned in Section 4.1.
 A distinction by  surface gravity
is made because systematic errors, especially non-LTE effects, may be
gravity-dependent; no obvious differences between the samples
are seen in the  Figures.
 No attempt has been made to adjust the plotted quantities  for
possible systematic differences between the various analyses.
Where a star has been analysed more than once, we elected to
adopt the results of the analysis considered the most reliable, but the
differences between investigators are small, and alternative methods
of handling the duplications would not lead to different
conclusions.

Figure 3 shows a well-known result (Spite \& Spite 1978):
 [Ba/Fe] $\simeq$ 0 for [Fe/H]
$\geq -2$ with a  decline to [Ba/Fe] $\simeq -1$ at 
[Fe/H] $\simeq -3$.
  For [Fe/H] $> -1$, extending the interval
 shown in Figure 3, 
 [Ba/Fe] $\simeq$ 0 (Edvardsson et al. 1993).
The scatter in [Ba/Fe] is particularly large for the  stars with [Fe/H] $< -2$.
(Stars obviously enriched in $s$-process products have been
excluded but some stars mildly enriched
 may remain undetected in the samples -- see
below.)
Viewed in Figure 3,  HD 140283 has a low [Ba/Fe] for its metallicity.
The star at [Fe/H] $= -1.7$ with the low [Ba/Fe] ($= -1.4$)  
is BD+80$^\circ$245,
a recognised $\alpha$-poor star (Carney et al. 1997; Stephens \&
Boesgaard 2002). HD 140283 is not
$\alpha$-poor (Magain 1989; Gratton \& Sneden 1994).

If the $r$-process relative yields of barium and europium are
solar-like, one expects [Ba/Eu] to approach $-0.7$ as the $r$-process
dominates the composition. In Figure 4, we show [Ba/Eu]
versus [Fe/H] for the same samples as plotted in Figure 3, except
that 
 a  Eu abundance was not always reported in the selected references. 
By inspection, there is a lower bound to the points  for [Fe/H] $< -2$ --
see the dashed line  at [Ba/Eu] $= -0.7$ (Figure 4), the solar $r$-process
limit. 
 Figure 4b shows that the lower bound to 
 [Ba/Eu] begins to rise at 
[Fe/H] $\simeq -1$  (see the solid line in Figure 4b) to match the solar
ratio ($\equiv$ 0) at [Fe/H] $\simeq$ 0 (Edvardsson et al. 1993;
Woolf, Tomkin, \& Lambert 1995; Koch \& Edvardsson 2002).\footnote{Two stars
from Fulbright's (2000) sample extend
the lower bound to about [Fe/H] $= -0.5$. Close  spectroscopic
scrutiny of this pair is warranted in order to confirm the apparent
dominance of the $r$-process at this high metallicity; Fulbright
measured only  Ba and Eu  among the set of heavier
elements.}

If Gratton \& Sneden's (1994)
estimate of [Ba/Eu] ($\simeq -0.7$) is adopted,  HD 140283 has the [Ba/Eu]
ratio expected of the $r$-process. If [Ba/Eu] $= -1.3$, as
Magain's (1989, also Magain \& Zhao 1990) analysis suggests,
HD 140283 has an unusually low [Ba/Eu] ratio. 
At our adopted [Ba/Eu] $= -1.05$, HD 140283 appears to fall 
below the lower bound for the $r$-process as set empirically
by the four heavily $r$-process enriched and other stars, but
improved measurements of Eu\,{\sc ii}  (and other) lines in HD 140283
would be welcomed.
 Note in Figure 4b that Magain's results for other stars 
tend to  [Ba/Eu] $\sim -0.7$, and, therefore,
his   result for HD 140283, a similar star to
the rest of his sample, may be correct.
Magain's (1989) measurement of [La/Eu] is consistent within the
uncertainties with the $r$-process expectation. No other
elements heavier than barium have been measured in HD 140283.
Given this evidence, we may consider it likely that the barium in
HD\,140283's atmosphere was produced via the $r$-process with
a solar $r$-like mix of the heavier elements.

\subsection{HD 140283: [Sr/Ba] and [Y/Ba] Ratios}

Equivalence between relative abundances of heavy elements in
metal-poor stars and the solar system $r$-process
abundances does not extend to elements lighter than barium.
In Figure 5, we show [Sr/Ba] and [Y/Ba] versus 
[Fe/H] for metal-poor stars. It is  known (McWilliam et al. 1995a, b;
McWilliam 1998) that Sr and Y abundances
span a much larger range than do the abundances of Ba and heavier
elements at a given [Fe/H]. Figure 5 shows clearly
(McWilliam 1998) that there is a lower bound and also suggests
that there is an upper envelope to [Sr/Ba] and
[Y/Ba]. For [Fe/H] $> -1$, the lower bound and the upper
envelope converge and tend 
to [Sr/Ba] = [Y/Ba] = 0.
The upper envelopes in Figure 5 appear to increase with
decreasing [Fe/H], but present data would allow for a levelling
off for [Fe/H] $< -3$.
 The quartet of stars very
 strongly enriched in  heavy elements sit at or close to the
lower bounds,  which are  slightly
below those expected by removing the {\it main} $s$-process (the contribution
from AGB stars) from the solar abundances: 
[Sr/Ba] $= -0.1$ and [Y/Ba] $= -0.4$. The small differences are likely
due to the fact that Sr and Y in the solar mix receive a contribution
from the {\it weak} $s$-process (the contribution from massive stars) which
is also reduced in the case of metal-poor stars.

\begin{figure*}
\centering
\includegraphics[width=6cm,angle=90]{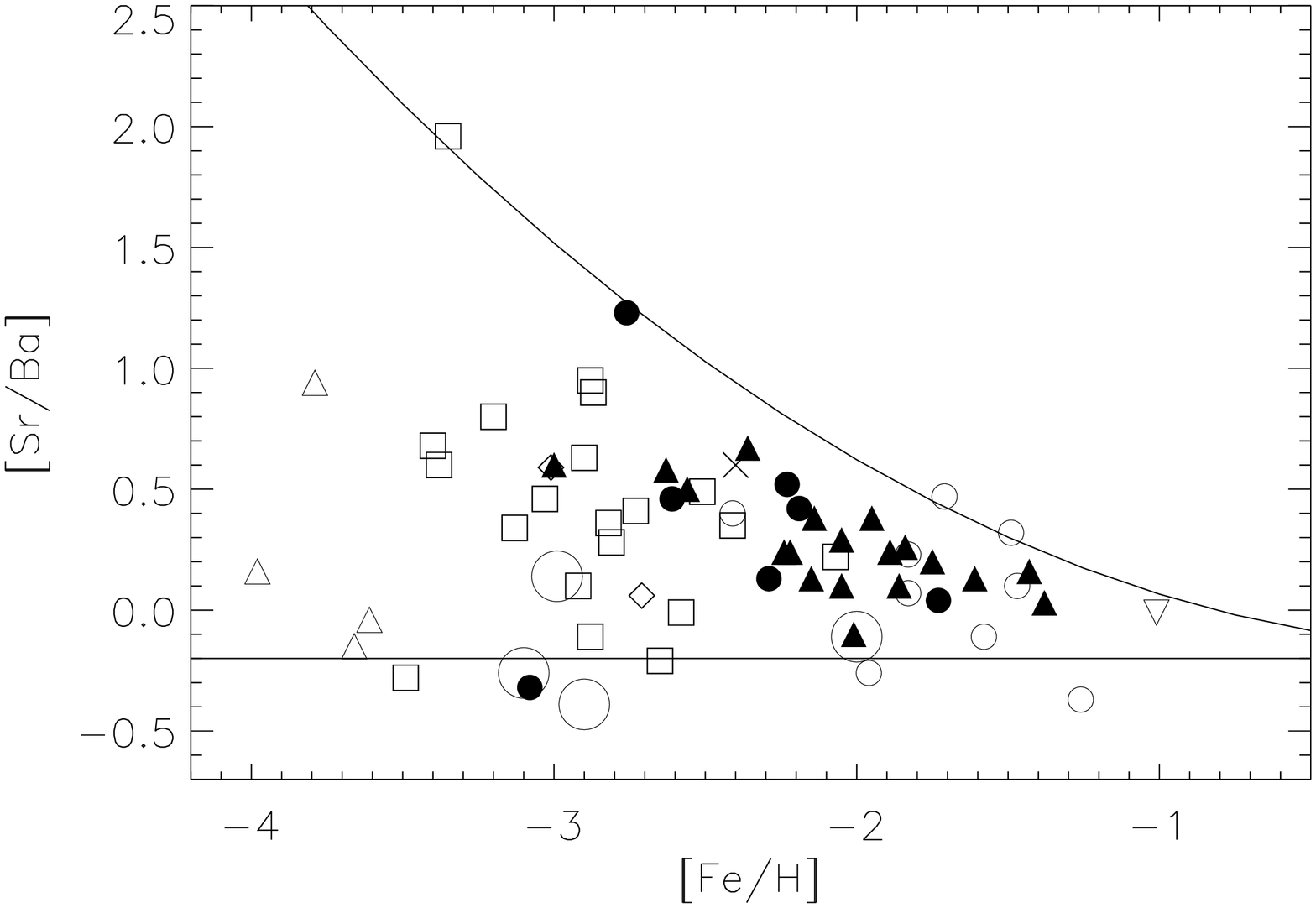}  
\includegraphics[width=6cm,angle=90]{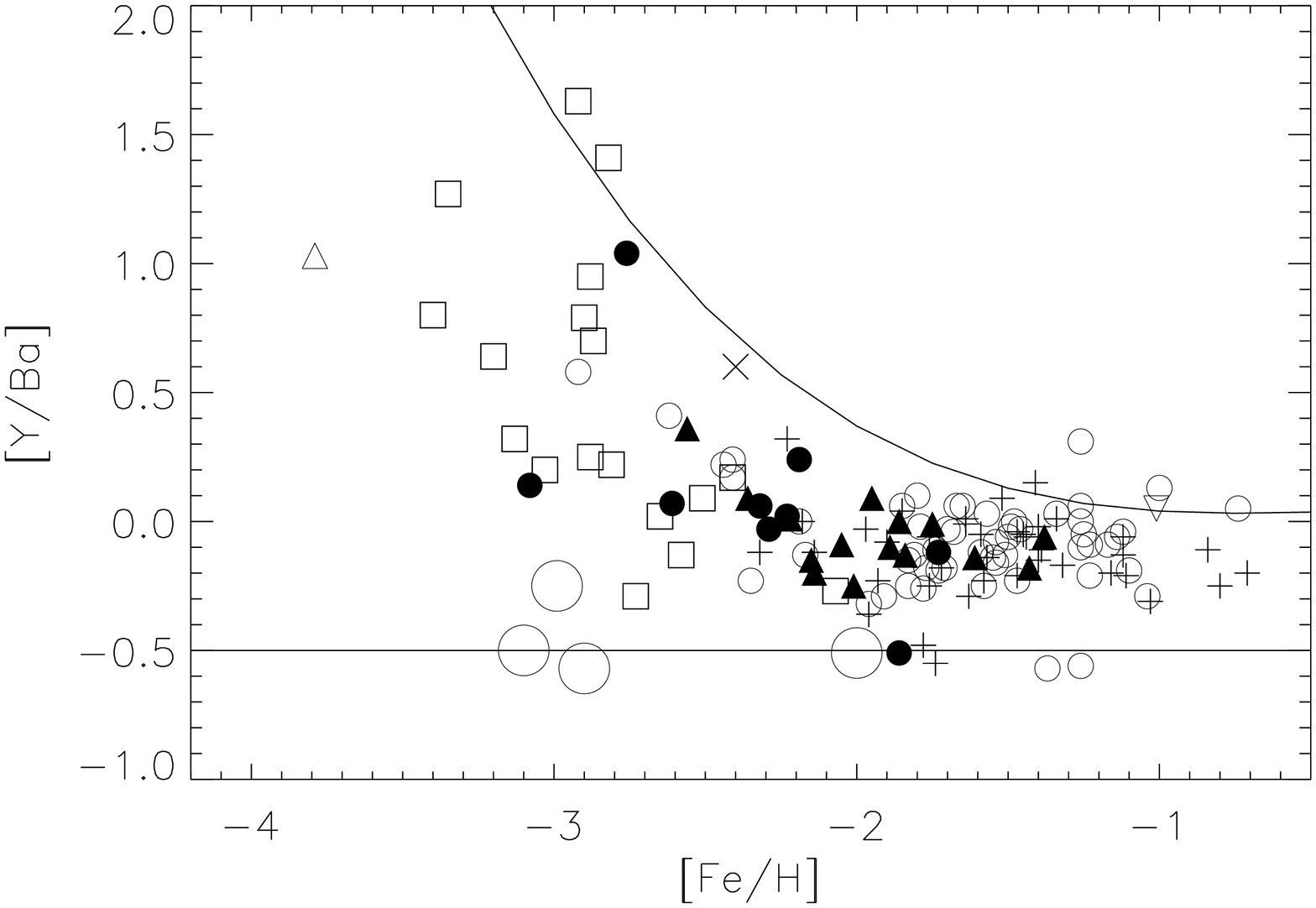}  
\protect\caption[ ]{
The abundance ratios [Sr/Ba] and [Y/Ba] versus [Fe/H] for representative
samples of metal-poor giants  and dwarfs. 
As before, HD 140283 is identified by the large cross,
the four severely $r$-process
enriched giants by the large open circles, and the other symbols are
explained in Figure 3. The solid lines sketch the possible boundaries to
the points.
\label{basry}}
\end{figure*} 

HD\,140283   sits in Figure 5 close to
the upper envelopes for its [Fe/H].   
The observation that many metal-poor stars 
 have [Sr/Ba] and [Y/Ba]
ratios in excess of the lower bound is part of the evidence
that  has led to suggestions that
there are multiple sites for the $r$-process (Wasserburg, Busso, \&
Gallino 1996; Qian \& Wasserburg 2001; Sneden et al. 2000).
In particular, Qian \& Wasserburg (2001) explore a scenario involving three
sources of heavy elements: (i) a pre-Galactic or `prompt' component
of low metallicity ([Fe/H] $\sim -3$) contributing Sr, Y, and Zr, (ii)
a common type of supernova providing heavy elements Sr to Ba to U in
solar $r$-process ratios with smaller relative abundances
for the 
elements lighter than barium, and (iii) a rarer type of supernova 
contributing iron, the lighter of the heavy elements, and possibly some barium. 
In such a composite picture, the stars with [Fe/H] $\sim -3$ were formed
from pre-Galactic gas to which a common supernova or two  added its
ejecta.  Qian \& Wasserburg's recipe matches HD 140283's abundances 
satisfactorily when ejecta of 3 common supernovae
 are mixed with the pre-Galactic
gas.
In this mix, the bulk of the Sr and Y are provided
by the prompt material, and the Ba and Eu   by the common
 supernovae.

Strontium and Y (also Zr)
abundances in HD 140283 show that  these lighter elements 
do not have the pattern of the solar $r$-process. In sharp contrast,
those stars most heavily enriched in $r$-process products do have
Sr and Y abundances (relative to Ba) that are quite similar
to solar $r$-process fractions.
Whether this difference between HD 140283 and the purest samples
of $r$-process products should affect the expected
barium isotopic mixture is unclear.

\subsection{The Barium Isotopic Ratio in other Metal-Poor Stars}

Magain's method of determining $f_{\rm odd}$ from the width of the
4554 \AA\ line has, so far as we are aware, not been applied to
other metal-poor stars. An alternative method  was
introduced by Magain \& Zhao (1993b) for stars somewhat more metal-rich
than HD 140283. Cowley \& Frey (1989) had earlier discussed this
approach to an isotopic analysis. 
  In this procedure, the Ba abundance
is derived from the 4554 \AA\ and also from at least one of
 the excited lines at 5853 \AA, 6142 \AA, and 6496 \AA. The 
abundance derived from an excited line in a metal-poor star  
is independent of the isotopic
and hyperfine splitting; the splittings are  too small to affect these lines.
 In appropriate stars, the much
stronger 4554 \AA\ line will be saturated, and, hence, the  equivalent width 
is sensitive to the isotopic and hyperfine splitting, that is to
 $f_{\rm odd}$. Then, $f_{\rm odd}$ follows by requiring that the
resonance line give the same total Ba abundance as the excited lines.

In an initial application to four stars with [Fe/H] $\sim -2$,
Magain \& Zhao (1993b) found a mean value for $f_{\rm odd}$ close to the
prediction from the solar $r$-process. Mashonkina \& Gehren (2000)
analysed about 20 stars  using their non-LTE calculations of
the Ba\,{\sc ii} lines. The abundance difference between the 4554 \AA\
line and excited lines was calculated for two isotopic mixtures:
the solar mix ($f_{\rm odd}$ = 0.18), and the solar $r$-process mix
($f_{\rm odd}$ = 0.46).
 For stars with  $-2 \leq$ [Fe/H] $\leq -0.5$ for which the
method is useful, the abundance from the 4554 \AA\ line was
generally more consistent
with that from the excited lines when the $r$-process  and not the
solar isotopic mix was assumed: the abundance difference was only about
0.1 dex between predictions for the two isotopic mixtures. 
 The [Ba/Eu] and [Sr/Ba] ratio for most of the stars are
close to the $r$-process limiting values: Mashonkina \& Gehren
give these ratios for some stars, and for others we obtain them from
Fulbright (2000). For such a sample, the $r$-process isotopic mix is 
not an unexpected result. 
Earlier, Mashonkina, Gehren, \&
Bikmaev (1999) applied the method to 
 a star with [Fe/H] $= -2.2$ of unknown [Ba/Eu] with
 a normal [Sr/Ba] ($= -0.18$) for which a solar mix
of the isotopes was found.  
This method is sensitive 
to atmospheric effects such as
the microturbulence, line broadening,  and  departures from LTE,
 as Cowley \& Frey (1989) had noted.

\subsection{Cosmic Scatter in the $s$ to $r$ Ratio for Metal-poor Stars?}

Although the scatter in [Ba/Eu] at a fixed (low) [Fe/H] is much less
than displayed by either [Ba/Fe]  or [Eu/Fe], there  appear
to be metal-poor stars with [Ba/Eu] above  the lower bound by more than
the supposed errors.
A simple way to give a star a high [Ba/Eu] for a given [Fe/H] is to
enrich it in $s$-process products.\footnote{The 
few known very metal-poor analogues of the Barium stars
and other examples of markedly $s$-process enriched stars have
been excluded from Figures 3 and 4.} 
The $s$-process enriched material may have been present in the natal clouds,
or accreted by passage through an interstellar cloud with a higher than average
abundance of $s$-process products.
Alternatively, the   
$s$-process products may have been  transferred from
a companion, as occurs in the classical Barium and CH stars.  
Contamination of some stars by $s$-process products was suggested
earlier in our discussion of Mg isotopic ratios in metal-poor stars
(Gay \& Lambert 2000).
(There is too the possibility that the He-core flash in
a low mass giant may trigger neutron release and synthesis of heavy
elements.  Such an event is unlikely to provide a solar-like
mix of the elements. If the core flash is responsible for the differences
in [Ba/Eu] ratios, giants but not dwarfs should show this effect. Figure 4
seems to contradict this idea.)

Natal clouds with differing compositions are thought to
be responsible for the spread in heavy $r$-process elemental 
 abundances  between
stars of a common very low [Fe/H]. As shown in Fig. 1, scatter decreases with
increasing [Fe/H] as the number of $r$-process contributors
grows. 
As long as the
relative yields of Ba and Eu from the $r$-process are the same
for all contributing  Type II supernovae, 
if the only production mechanism is the $r-$process, [Ba/Eu] will show no
scatter. That could be the case in Fig. 4 for the lowest [Fe/H].
Donors of $s$-products are surely more slowly evolving than
the supernovae contributing $r$-process products. Clouds
first contaminated with $s$-products may form stars of similar [Fe/H]
but differing [Ba/Eu]. Onset of this scatter is expected at a [Fe/H]
greater than that marking the onset of $r$-process scatter, i.e., the
minimum observed [Fe/H] ($\simeq -4$). This speculation is consistent 
with the data
plotted in Figure 4.

If our suggestion about  $s$-process  contamination 
causing the scatter in [Ba/Eu] is correct, other  
heavy
elements with an $s$-process contribution should  show a scatter correlated with the
enhancement of [Ba/Eu] above the lower bound.
This idea is tested in Figure 7 using Johnson \& Bolte's data.
 Figure 7 appears to show that stars with [Ba/Eu] in excess of the
$r$-process lower bound are also overabundant in other (s-process dominated)
 heavy elements.
 We note
that the trail of
data points in the various panels commences from about the expected
and well-populated endpoint for the solar $r$-process ratios. 
The apparent trends seem to favour our hypothesis.
Unfortunately, a dysprosium
abundance was provided by Johnson \& Bolte for just one of the six
`Ba-rich' ([Ba/Eu] $> -0.3$) stars. Burris et al. give Dy abundances for
an additional two `Ba-rich' stars which at [Dy/Eu] $= -0.07$ and 0.11
indicate, as required by our speculation, that Dy is not enriched
in the `Ba-rich' stars. However, we must recognize that the Dy abundances
of the stars with [Ba/Eu] at the $r$-process value show a large spread
presumably
reflecting measurement errors for this
element. 
  Perhaps, our suggestion
about the onset of the $s$-process as a contributor to
the scatter in relative abundances will stimulate the pursuit of
a higher accuracy in the abundance determinations of these
heavy elements.

\begin{figure*}
\centering
\includegraphics[width=10cm,angle=90]{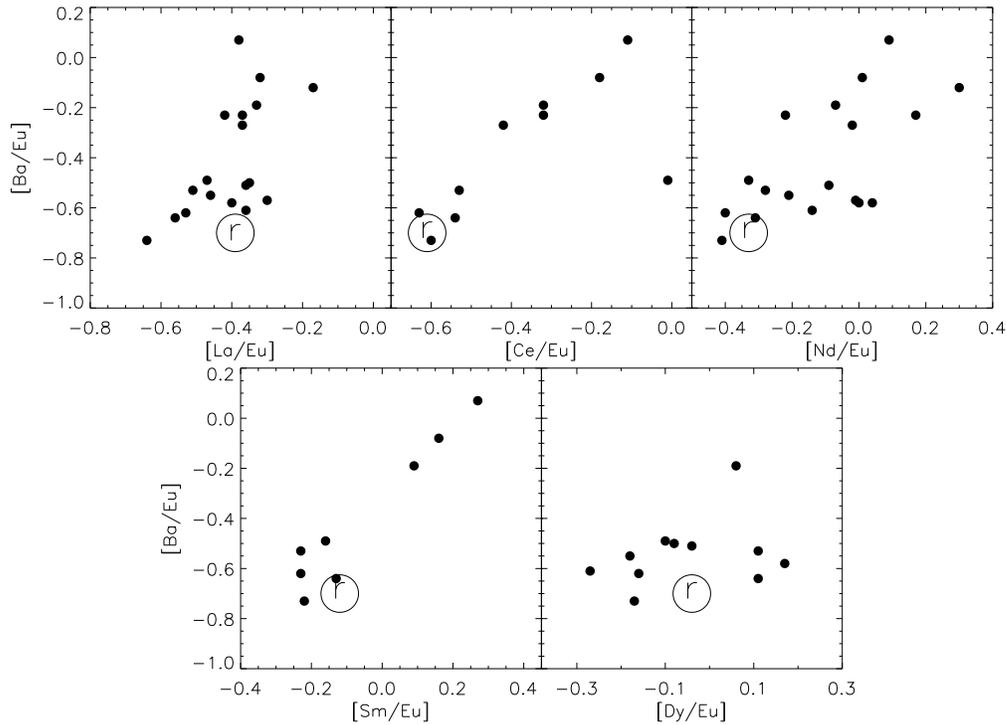}  
\protect\caption[ ]{
The abundance ratio [Ba/Eu] versus [X/Eu], where X 
is La, Ce, Nd, Sm, and Dy, for giants analysed by Johnson
\& Bolte (2001). The solar $r$-process ratio is indicated in
each panel by the circled-$r$. 
\label{smix}}
\end{figure*}

\section{Concluding Remarks}

Our profile analysis of the Ba\,{\sc ii} 4554 \AA\ line does not confirm
Magain's (1995) result  that the mix of odd to even isotopes,
$f_{\rm odd}$, is  not the high value ($\simeq$ 0.5) expected of
the $r$-process on the assumption that the $r$-process always 
provides a  solar $r$-process mix of heavy nuclides.
On the contrary, our estimate of $f_{\rm odd}$ is compatible with
a solar $r$-process mix of the barium isotopes.

 Magain argued also that the then available data on the
abundances of heavy elements in  metal-poor
stars did not support the  view that the $r$-process
contributions were dominant, and, thus, his measurement of
a low value for $f_{\rm odd}$ was not at odds with
his view of the pattern of elemental
abundances. In striking contrast, the evidence now available
shows that the abundances for heavy elements in metal-poor stars are in general
reflective of an $r$-process with yields for barium and heavier
elements remarkably similar to the solar case.
Although a thorough analysis of the heavier elements present in
HD 140283's atmosphere has yet to be presented, it seems 
likely that a solar-like $r$-process mix is present.
Then, it  is not  surprising that  the barium isotopic mix
is similar to that of the solar $r$-process fraction of barium.

Extensive and accurate work on the
neutron capture cross-sections for barium isotopes has enabled
$f^s_{\rm odd}$ and, hence, $f^r_{\rm odd}$
  to be determined accurately (Arlandini et al.
1999). 
More directly,  analyses of the stars with extreme
enrichments of the $r$-process products give a Ba/Eu ratio  consistent
with the estimated solar $r$-process fractions.
That the solar $r$-process pattern of elemental
abundances fits  the abundances of barium and heavier
elements in metal-poor stars is a remarkable
empirical fact.
 Sneden et al. (2002) have shown that the
europium isotopic abundances in the heavily $r$-process enriched metal-poor
stars have their solar $r$-process ratio. 
There remains the observation that the Sr/Ba and
Y/Ba ratios of metal-poor stars often exceed that expected
of the $r$-process and observed in those heavily $r$-process
enriched stars. As Figure 5  shows, HD 140283 seems to be
among the stars with the greatest Sr/Ba and Y/Ba ratios.
For this and similar stars, the Sr and Y
may have been provided by the pre-Galactic gas (Qian \& Wasserburg 2001).
Can this material harbor distinct isotopic mixtures for barium and other
elements, especially Sr, Y, and Zr?

Our analysis clearly shows that the barium isotopic mixture has
a rather subtle effect on the 4554 \AA\ profile. High-resolution
spectra of high S/N ratio are just one prerequisite for a determination of
$f_{\rm odd}$. 
With today's large telescopes and the development of
hydrodynamical model  atmospheres,
 it may be possible to refine and extend the
isotopic analyses to other metal-poor stars and to other elements.

\section{Acknowledgments}

We are indebted to Ram\'on J. Garc\'{\i}a L\'opez for obtaining 
part of the observations
used in this work. We thank Chris Sneden and John Cowan for 
helpful discussions. NSO/Kitt Peak FTS data used here were produced by 
NSF/NOAO.
This research was supported in part by the
National Science Foundation (grant AST-0086321) and the Robert A. Welch
Foundation of Houston, Texas (grant F-634).

\end{document}